\documentclass{aa}  

\usepackage{graphicx}
\usepackage{tablefootnote}
\usepackage{threeparttable}
\usepackage{txfonts}
\begin{document}

   \title{Exploring the region encompassing $\gamma$\,Cygni SNR and MAGIC\,J2019$+$408 with the GMRT at 325 and 610 MHz}
   
   \titlerunning{Exploring the region encompassing $\gamma$\,Cygni SNR and MAGIC\,J2019$+$408}


   \author{J.M. Paredes
          \inst{1}
          \and
          P. Benaglia\inst{2}
          \and
          C.H. Ishwara-Chandra\inst{3}
          \and
          V. Bosch-Ramon\inst{1}
          \and
          Marcel Strzys\inst{4}
          }
          
\authorrunning{J.M. Paredes}

   \institute{Departament de Física Quàntica i Astrofísica, Institut de Ciències del Cosmos, Universitat de Barcelona, IEEC-UB, Martí i  Franquès 1,
08028, Barcelona, Spain\\
              \email{jmparedes@ub.edu, vbosch@fqa.ub.edu}
         \and
             Instituto Argentino de Radioastronomía, CONICET-CICPBA-UNLP, CC5 (1897) Villa Elisa, Prov. de Buenos Aires, Argentina\\
             \email{paula@iar-conicet.gov.ar}
        \and
             National Centre for Radio Astrophysics, Tata Institute of Fundamental Research, Pune University Campus, Pune, 411007, India\\
              \email{ishwar@ncra.tifr.res.in}
        \and
         ICRR, The University of Tokyo, 277-8582 Chiba, Japan\\
         \email{strzys@icrr.u-tokyo.ac.jp}
         }
         
   \date{Received ; accepted }

 
  \abstract
   {$\gamma$\,Cygni is a young supernova remnant located in the Cygnus region. MAGIC (Major Atmospheric Gamma Imaging Cherenkov) telescopes detected TeV emission (MAGIC\,J2019$+$408) to the north-west of this remnant, $\sim 5 \arcmin$ from its border.}
   {We want to identify the radio sources within the region encompassing $\gamma$\,Cygni and MAGIC\,J2019$+$408 to shed light on their nature and investigate if these radio sources could be potential contributors to gamma-ray emission.}
   {We carried out a detailed study of the data we obtained with a survey of the Cygnus region at 325 and 610 MHz with the Giant Metrewave Radio Telescope (GMRT).}
   {We detected several radio sources in the region where the radio and the TeV emission overlap, as well as several areas of enhanced radio emission. In particular, two of these areas of diffuse enhanced emission may correspond to the supernova remnant interacting with a high density region, which seems to be the best candidate for the MAGIC source. Another two radio sources, which may or may not contribute to the gamma rays, are also spatially coincident with the emission peak of the MAGIC TeV source. 
   One of them displays a rather peculiar extended morphology whose nature is completely unknown.
   }
   {We have identified the radio sources overlapping $\gamma$\,Cygni and MAGIC\,J2019$+$408 and have shown that their potential gamma-ray contribution is likely not dominant. In addition, some of the studied sources show peculiar physical characteristics that deserve deeper multi-wavelength observations.}


   \maketitle
%

\section{Introduction}

   The $\gamma$\,Cygni SNR (or G78.2+2.1) is a young ($\sim 7\times10{^3}$ years) supernova remnant (SNR)  located in the Cygnus region at a distance of $\sim 1.8$~kpc \citep{Higgs1977, Uchiyama2002}. Aperture-synthesis observations at 1.4 GHz showed that the $\gamma$\,Cygni SNR has a shell structure of  $\sim 62 \arcmin$ in diameter with a non-uniform surface brightness, being the radio emission strongest in the south-east quadrant, whereas a second region of enhanced non-thermal radio emission is present in the north-west quadrant \citep{Higgs1977}. Further studies showed an  averaged SNR spectral index of $\alpha= -0.75 \pm 0.03$ ($S \propto \nu^\alpha$), with spatial variations ranging from $-0.40$ to $-0.80$ \citep{Zhang1997, Ladouceur2008}. The spectrum was harder in the west and north-west part of the SNR, while the softest index was found in the south part. 

   Observations by  {\it ASCA} showed that in addition to a thermal component explaining the overall X-ray flux, there is a hard X-ray component from several clumps in the northern part of the remnant \citep{Uchiyama2002}. Later on, observations by {\it Chandra} -- covering significant parts to the north and north-west and central parts of the $\gamma$\,Cygni SNR -- resolved these clumps, separating the emission coming from the compact objects and the emission coming from the surrounding X-ray diffuse emission \citep{Leahy2013}.

   At GeV energies, {\it Fermi}-LAT detected extended emission towards the source \citep{Lande2012}, as well as a bright excess in the north-west region with a spectral index harder than in other parts of the shell \citep{Fraija2016}. Near the centre of the SNR, {\it Fermi}-LAT detected the $\gamma$-ray pulsar PSR~J2021+4026, with a spin-down age of 77 kyr \citep{Abdo2009} and no radio pulsar counterpart associated with it. The gamma-ray instrument {\it AGILE} found that the {\it AGILE}-GRID off-pulsed emission profile above 400 MeV appears to partially cover the south-east rim of the shell, indicating that part of this emission comes from the shell \citep{Piano2019}.
   
   At TeV energies, VERITAS discovered a spatially extended source, VER\,J2019$+$407, located on the north-west rim of the $\gamma$\,Cygni SNR \citep{Aliu2013}. Recently, by combining  9 years of {\it Fermi}-LAT data and 87 hours of good-quality MAGIC data, three gamma-ray emission regions have been identified: the SNR interior, an extended emission located immediately outside the SNR, `the arc', and a Gaussian-shaped extended source dubbed MAGIC\,J2019$+$408 to the north-west of the remnant \citep{MAGIC2020}. The TeV emission outside the shell has been interpreted as the result of cosmic rays (CRs) escaping the shock of the SNR upstream into the interstellar medium (ISM). For MAGIC\,J2019$+$408, this would require the presence of a region with a denser ISM. While the atomic and molecular material in the region provides a sufficient density, a clear counterpart could not be identified and thus the possibility for an extended source unrelated to the SNR cannot be ruled out \citep{MAGIC2020}.
   
  \citet{Benaglia2020, Benaglia2021}, using the Giant Metrewave Radio Telescope (GMRT) at 325 MHz and 610 MHz, performed a deep survey of the Cygnus region that included the $\gamma$\,Cygni SNR and its nearby surroundings. A detailed study of the $\gamma$\,Cygni SNR was beyond the scope of these publications and was planned to be carried out in a future work. In the present paper, we present such a detailed exploration at these frequencies of the region encompassing MAGIC\,J2019$+$408 as well as other sources in the field of the $\gamma$\,Cygni SNR.
   
   


\section{The 325 and 610~MHz radio data}
We have used the data obtained by \citet{Benaglia2020} in the GMRT 325 and 610 MHz Cygnus survey to study the radio emission in the field of the $\gamma$\,Cygni SNR, in detail, and the radio emission surrounding this field, namely towards MAGIC\,J2019$+$408. The radio observations were carried out during four campaigns between 2013 and 2017. The field of views (FoV) of GMRT are 81$\pm$4$'$ and 43$\pm$3$'$ at 325~MHz and 610~MHz bands, respectively. The synthesised beams of the final mosaics were $10'' \times 10''$ for the 325 MHz band, and $6'' \times 6''$ at the 610 MHz band. The attained rms values resulted, on average, in up to a 0.5~mJy~beam$^{-1}$ and a 0.2~mJy~beam$^{-1}$, respectively. Full details of the observational campaigns, data reduction, and mosaicked image construction at the two bands are given in \citet{Benaglia2020}. The 325~MHz radio image of the $\gamma$\,Cygni SNR region is shown in Fig.~\ref{RadioSNR}. The image shows different regions of enhanced diffused emission: to the north-west, to the south-east, and outside of the SNR. There are also several strong sources that we have identified with a numeric label and that we describe here in detail. The flux densities and the derived spectral indices $\alpha$ ($S \propto \nu^\alpha$) of these sources are summarised in Table~\ref{tab:results}. In the case of extended sources, we derived the spectral indices using the 610 MHz image convolved to the 325 MHz synthesised beam of  $10^{\prime\prime}\times 10^{\prime\prime}$. Other sources, mainly point-like sources, were identified and catalogued by \citet{Benaglia2020, Benaglia2021} and they are not considered here. 

\section{Radio emission towards MAGIC\,J2019$+$408}

\begin{figure}
   \centering
   \includegraphics[width=9cm, angle=0]{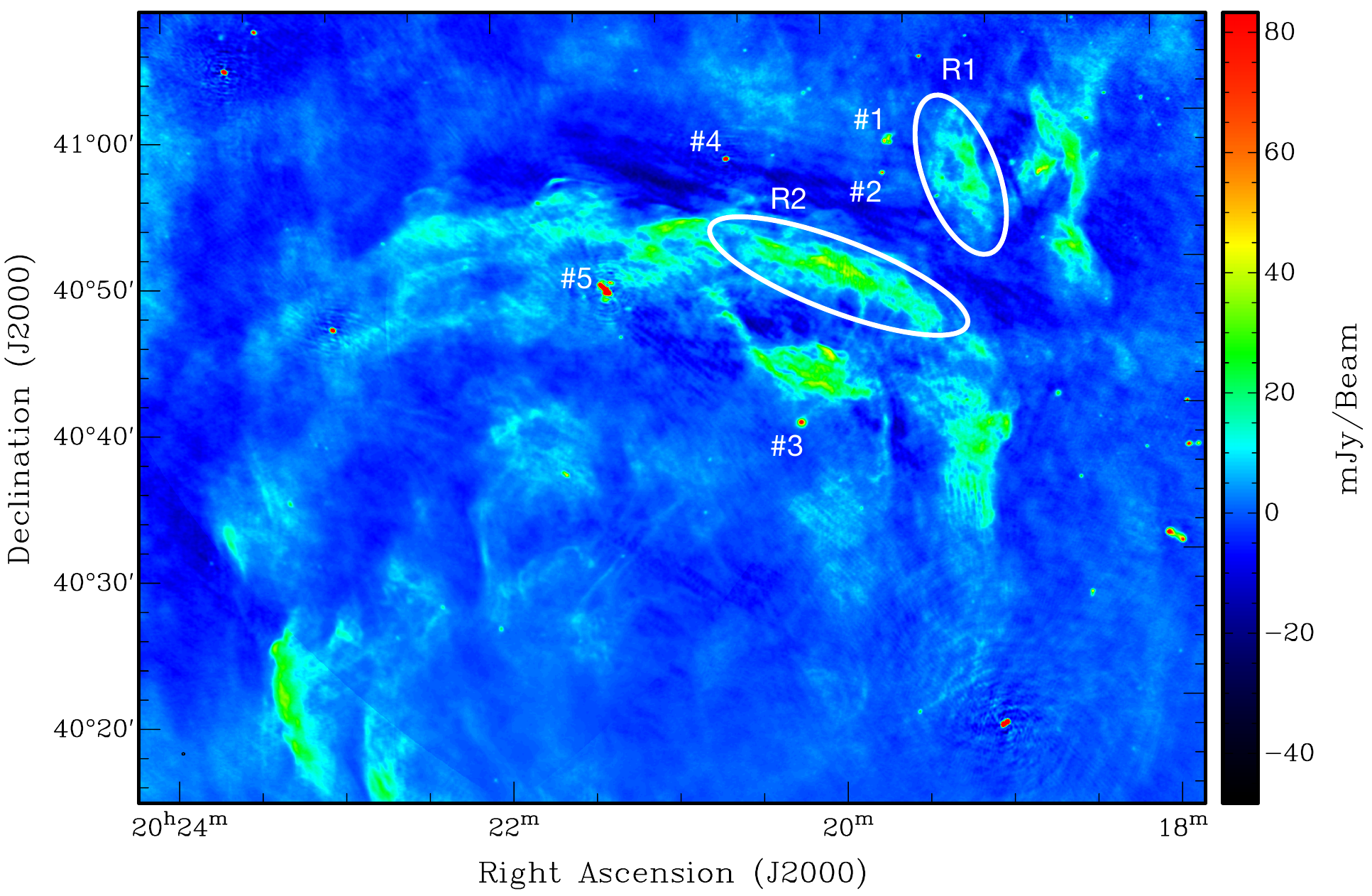}
   \caption{325 MHz radio image of the Cygni SNR (G78.2+2.1). The numeric labels, and R1 and R2, identify the sources and the diffuse enhanced emission regions,  respectively, that are discussed in detail here (see Table~\ref{tab:results}). The colour scale interval shown is ($-48,+83$) mJy beam$^{-1}$ to outline weaker features. The synthesised beam (11.95$''$~$\times$~9.61$''$, $-76.6^{\circ}$) is plotted at the bottom left corner. The rms, where the emission is lowest, is $\simeq$0.5~mJy~beam$^{-1}$}.
         \label{RadioSNR}
   \end{figure}

\begin{figure}
   \centering
\includegraphics[width=8.5cm, angle=0]{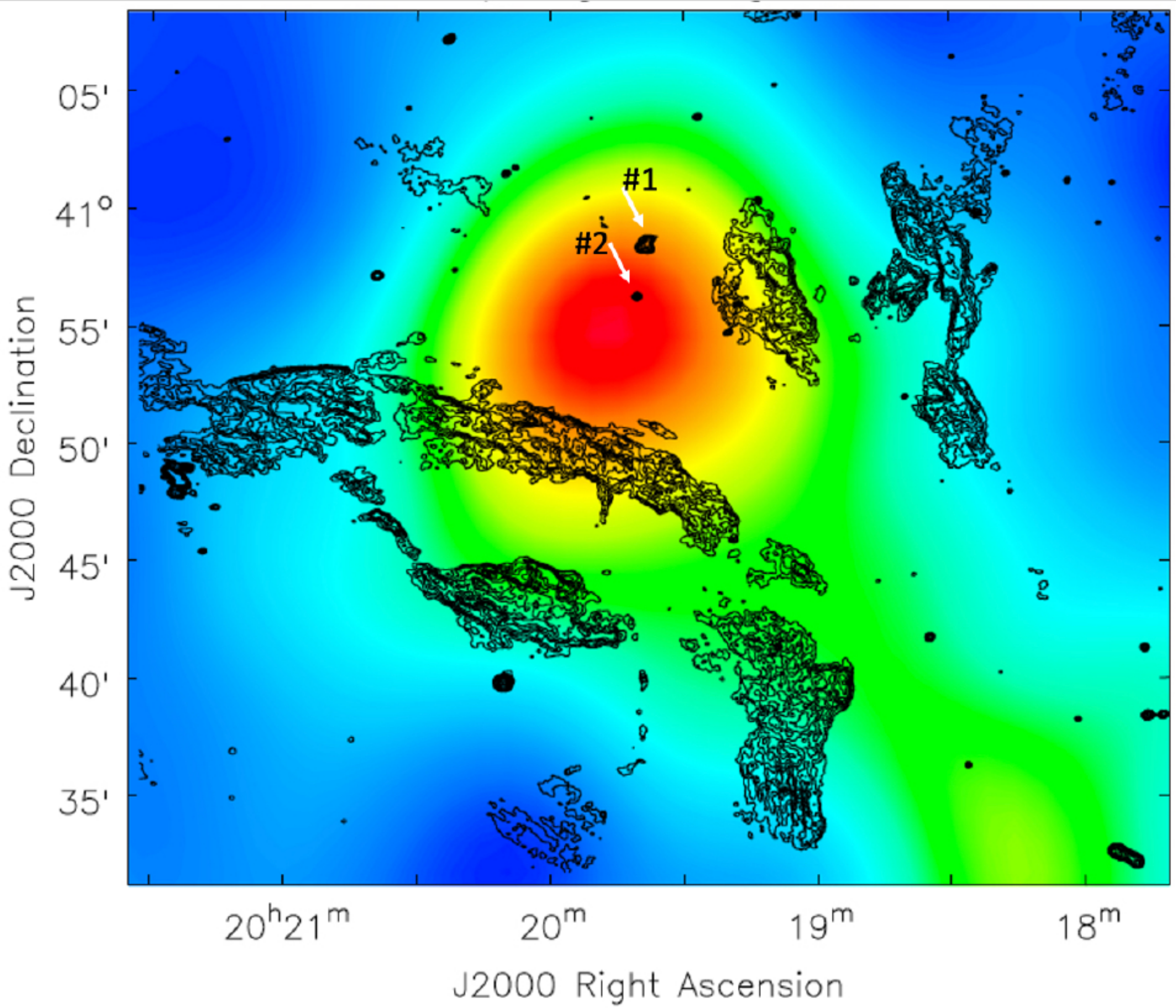}
      \caption{\footnotesize Sky map in units of relative flux (excess over background) of
the $\gamma$\,Cygni SNR region as observed by MAGIC at energies $E>250$ GeV  \citep[see][for details]{MAGIC2020}. The overlaid black contours represent the 325~MHz emission measured by GMRT. The sources Roman Squid (\#1) and G78.4+2.6 (\#2) are marked with white arrows (extended sources R1 and R2 at 325~MHz -- see Fig.~\ref{RadioSNR} -- lie to their west and south, respectively).
              }
         \label{MAGIC-SNR}
   \end{figure}

Our exploration of the region overlapping MAGIC\,J2019$+$408 ($E>250$~GeV) using GMRT 325~MHz data has revealed two regions (R1 and R2, see Fig.~\ref{RadioSNR}) of diffuse enhanced emission, and two radio sources (\#1 and \#2) spatially coincident with the peak of the MAGIC source (see Fig.~\ref{MAGIC-SNR}). One of these sources (\#2) is G78.4+2.6, a known ultracompact HII region displaying a cometary morphology at an estimated distance of 1.7~kpc \citep[see][and references therein]{Neria2010} or 3.3 kpc \citep{Kurtz1994}. The other source (\#1) displays a rather peculiar extended morphology, reminiscent of a {roman squid}, whose nature is completely unknown. An association of sources \#1 and \#2 with MAGIC\,J2019$+$408 is unlikely given that the TeV emission is clearly extended, whereas the radio sources should appear as point sources for MAGIC, particularly if they are of extragalactic origin.


\subsection{\textit{Roman squid} radio source (\#1)}

\begin{figure*}
\centerline{
 \includegraphics[width=9cm, angle=0]{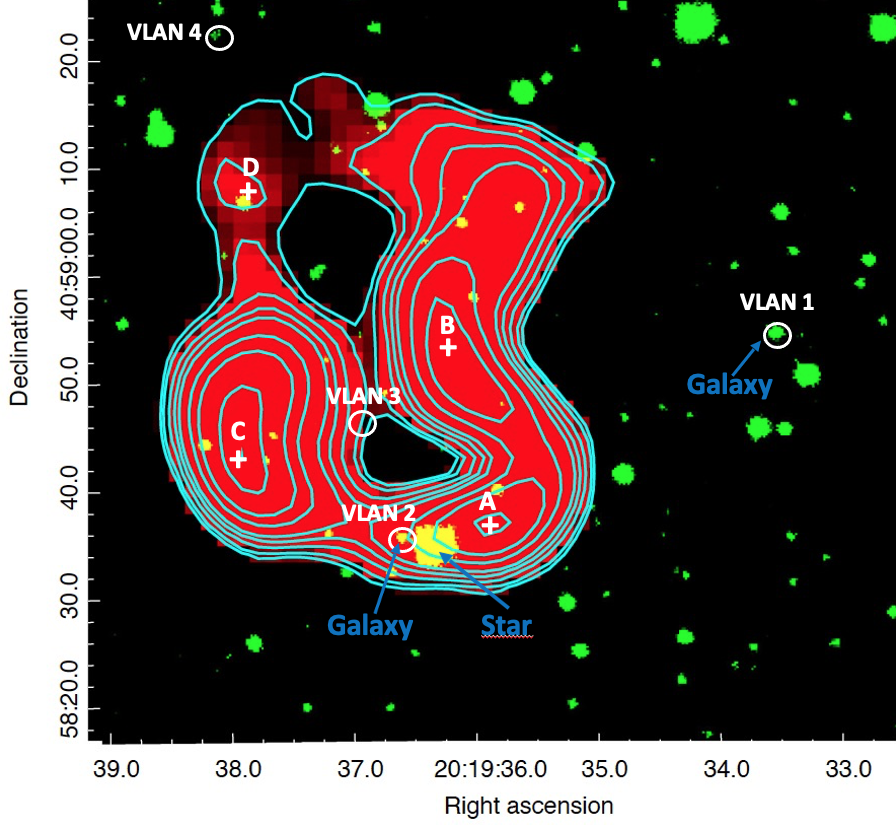}
 \includegraphics[width=9cm, angle=0]{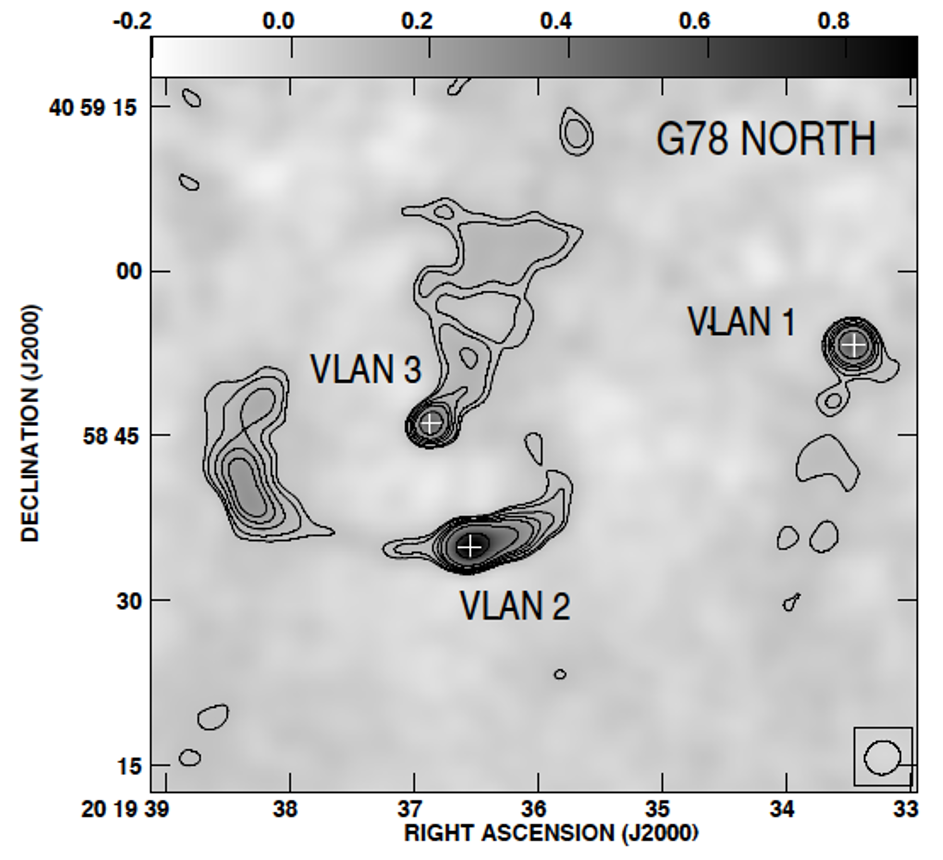}
}
\caption{\footnotesize  Radio images of the Roman Squid source. $\mathbf{Left}$: GMRT 610~MHz radio image of the Roman Squid source in red. The peak position of each of the four components A, B, C, and D are marked with a white cross. For clarity, we also show the contour levels, 3, 4, 6, 8, 10, 15, 20, 30, 35, and 40 times the rms noise (130~$\mu$Jy~beam$^{-1}$).
In green we show an infrared {\it K}-band image of the region from the UKIDSS catalogue. We note that the green  within the red region appears as yellow. The white circles indicate the position of the VLAN sources. $\mathbf{Right}$: VLA 8.4~GHz radio image of the  same field \citep{Neria2010}. 
Contour levels are 3, 4, 6, 8, 10, 15, 30, and 50 times the average rms noise of the image (21.8~$\mu$Jy~beam$^{-1}$)}.
 \label{FigVLAN}
\end{figure*}

The 610~MHz GMRT image shows an extended source of $23^{\prime\prime}\times 49^{\prime\prime}$ with a deformed ellipsoidal morphology consisting of several knots, designated as A, B, C, and D, with a fainter diffuse emission connecting them (see Fig.~\ref{FigVLAN}, left and Table~\ref{tab:results}). This structure, which we hereafter call Roman Squid, is also present in the 325~MHz image (not shown here), although with a somewhat lower resolution. The 325 and 610~MHz images suggest a physical connection between the knots (sources) that make up the structure. The values of the spectral indices of these sources are around $-1$, indicating a non-thermal origin of the emission and suggesting a common origin. 

\citet{Neria2010} reported the detection of a cluster of compact radio sources in the region, named VLAN\,1 to VLAN\,5 (the last one is located outside the shown images), from VLA 8.4 GHz observations (see Fig.~\ref{FigVLAN}, right). Sources VLAN\,2 and VLAN\,3, with a flux density of $0.87\pm0.04$ mJy and $0.49\pm0.04$ mJy, respectively, and co-spatial with the Roman Squid, show a core and a jet-like extension. Their search for known counterparts of the VLAN sources was inconclusive, with no optical or X-ray candidates in the surroundings within the immediate few arcseconds. The morphology of three independent sources, VLAN\,1 to 3, prompted \citet{Neria2010} to suggest a cluster of radio galaxies or alternatively HII regions surrounding massive stars.

The global morphology of the Roman Squid is not visible in the high resolution image at 8.4~GHz. In addition, when comparing our 610~MHz image with the 8.4~GHz image, we see that there is an offset of about 7$^{\prime\prime}$ (0.06 pc at a distance of 1.7 kpc) between the core position of VLAN\,2 and the peak emission of A, as well as between VLAN\,3 and the peak emission of B. However, it is very likely that these pairs of sources are associated with each other since the offset can be easily explained if the cores of VLAN\,2 and VLAN\,3 -- with flux densities $\lesssim$ 0.87 and 0.49 mJy, respectively -- are optically thick at 8.4 GHz, implying that the core counterparts are very faint at lower frequencies. On the other hand, the jet, which is optically thin at 8.4 GHz, can be stronger at lower frequencies. This can give rise to an offset in the centroid of the emission at low and high frequencies. We note that VLAN~3 might not be related to source B, in which case VLAN~3 would be an unrelated source located by chance in a way that resembles the core of a more extended structure.

We have been looking for possible counterparts at other wavelengths. At the infrared band, and using the UKIDSS  
Galactic Plane Survey Catalogue  \citep{Lucas2008}, we have found a star (UGPS J201936.28+405833.9) and a galaxy (UGPS J201936.53+405834.8) lying $\sim$5$^{\prime\prime}$ east of source A (see Fig.~\ref{FigVLAN}, left). The star, with coordinates  RA$_{\rm J2000}$= 20h19m36.2813s and DEC$_{\rm J2000}$= +40$^{\circ}58'33.946"$ and a magnitude of $K$ = 11.17, is placed at a distance of 1.8 kpc (Gaia EDR3) and can be classified as a B5 V or A0 V star with $A_{\rm V} \sim 5$ according to the photometry of 2MASS. The galaxy is situated at RA$_{\rm J2000}$= 20h 19m 36.5347s and DEC$_{\rm J2000}$= +40$^{\circ}58'34.806''$ and has a magnitude of $K$ = 16.16. This galaxy is positionally coincident with VLAN\,2 (RA$_{\rm J2000}$= 20h19m36.554s and DEC$_{\rm J2000}$= +40$^{\circ}58'34.86''$). Also, VLAN\,1 (RA$_{\rm J2000}$= 20h19m33.457s and DEC$_{\rm J2000}$= +40$^{\circ}58'53.25''$) is positionally coincident with UGPS J201933.46+405853.0, catalogued as a galaxy with a probability $\geq$ 90\%.

In the X-ray band, {\it ROSAT} observed the area around the \textit{Roman Squid} obtaining only a poorly constraining upper limit because of its relatively low sensitivity and angular resolution. The satellite {\it Chandra} also observed this area, but unfortunately these observations did not include the region encompassing the Roman Squid. The absence of any X-ray counterpart however prevents any assessment on the galactic or extragalactic nature of the Roman Squid being made through the estimation of the distance.

The identification of two galaxies associated with VLAN\,1 and VLAN\,2 gives support to the suggestion given by \citet{Neria2010} that VLAN\,1 to 3 are members of a cluster of radio galaxies, although, in the case of VLAN\,3, no infrared counterpart has been found.

Physical scenarios that could roughly explain the morphology of the Roman Squid and the absence of astrophysical objects behind B, C, and D may include head-tail radio galaxies \citep{Sebastian2017}. This type of radio galaxies occur in clusters of galaxies and are characterised by a head identified with the optical galaxy and two tails sweeping back from the
head, forming an angle with the galaxy at the apex. These sources are understood to be Fanaroff–Riley type\,I \citep{Fanaroff1974} radio sources moving through the gas in the cluster, and the shape of the source is due to the diffuse radio emitting plasma being deflected and decelerated by the intracluster medium.

The presence of a galaxy just at the position where it would be expected in the case that the Roman Squid were the head of a head-tail radio galaxy supports the extragalactic scenario. The morphology and size of the Roman Squid resembles the head of the  head-tail radio galaxy NGC\,1265 that has a tail extending to several arcmins \citep{Sebastian2017}. Deeper radio observations of the Roman Squid are necessary to evaluate the head-tail radio galaxy scenario as a possible explanation. A galactic scenario would also be possible if the radio emission was coming from a runaway pulsar wind nebula (PWN), which could produce bent jet-like features radiating in the X-ray band  \citep[see e.g.][and references therein]{Kargaltsev2017}, explaining the distinct morphology observed in the Roman Squid. If the star (UGPS J201936.28+405833.9) was associated to VLAN\,2 and source A, this would also favour a galactic location for the source, most likely of a binary nature. In that context, a compact object hosted by the binary could produce outflows that would be affected by proper motion, also explaining the shape of the Roman Squid \citep[e.g.][]{yoon2011,bb11,Bosch-Ramon2011}. In all these scenarios, VLAN\,3 would be a background source, as noted above.

It is worth noting that the interpretation that the TeV emission is the result of CRs escaping the shock of the SNR upstream into the ISM is robust and leaves a possible GeV-TeV contribution associated to the Roman Squid, or to parts of it,
 at a negligible level. This possible contribution from the Roman Squid, if any,
 could come from a counterpart associated to a radio galaxy, a PWN, or a binary system. These types of source have been shown to be gamma-ray emitters. The detection of some gamma-ray variability, linked to emission coming from the source core in the case of a radio galaxy or a binary system, would give support to these scenarios.

\begin{table*}
\caption{\footnotesize Flux density and spectral index of GMRT radio sources detected in the region where the radio and the TeV emission overlap. }
\begin{threeparttable}[t]
\centering
        \begin{tabular}{c c c c c c c c}
         \hline

{ID} & {Source} & RA$_{\rm J2000}$ & DEC$_{\rm J2000}$ & {$S_{\rm 325MHz}$}  & {$S_{\rm 610MHz}$} & {Spectral index $\alpha$} & Remarks \\  
     & {name} & {(hh:mm:ss)} & {(dd:mm:ss)} & (mJy) & (mJy) & ($S \propto \nu^\alpha$) & \\
\hline

\#1 & Roman Squid$^{\rm a}$ &20:19:36.83 &40:58:54.61 & $98.1\pm1.0$ & $49.0\pm0.7$ &$-1.1\pm0.1$ & \\
 & A$^{\rm b}$  & 20:19:35.99 &40:58:36.62 & $17.3\pm1.5$ & $9.2\pm0.7$ &$-1.0\pm0.2$  & VLAN 2 \\
  & B$^{\rm b}$  & 20:19:35.98 &40:58:55.21 & $48.0\pm2.7$ & $22.9\pm1.6$ &$-1.2\pm0.1$ & VLAN 3 \\
   & C$^{\rm b}$  & 20:19:37.76 & 40:58:45.68 & $40.5\pm1.8$ & $19.0\pm1.0$ &$-1.2\pm0.1$ & \\
   & D  & 20:19:37.97 & 40:59:08.95 & $6.2\pm0.3$ & $4.0\pm0.3$ &$-0.7\pm0.1$  & \\
\#2 & G78.4+2.6$^{\rm b, c}$ & 20:19:38.89 &40:56:36.21 & $36.5\pm1.4$ & $63.9\pm0.8$ &$0.9\pm0.1$ & VLA, IR \\
\#3 & Clump C2$^{\rm b, d}$ & 20:20:11.44 &40:39:39.12 & $403.9\pm12.4$  &  $167.1\pm3.2$ & $-1.4\pm0.1$ & X-ray\\ 
\#4 & YSO$^{\rm b}$ &20:20:35.65 &40:57:54.84 & $189.8\pm1.3$ & $227.0\pm0.4$ &$0.3\pm0.1$ & X-ray\\
\#5 & Clump C1$^{\rm d}$ &  & &  & &  & X-ray \\
 & C1-W$^{\rm b}$  &20:21:18.89  &40:49:39.94 & $83.6\pm1.3$ & $43.7\pm1.2$ &  $-1.0\pm0.1$  &  \\
  & C1-MN$^{\rm b}$  & 20:21:20.57 &40:49:14.98 & $939.0\pm2.4$ & $414.5\pm1.0$ &  $-1.3\pm0.1$ &  \\
  & C1-MS$^{\rm b}$  & 20:21:19.75 &40:48:59.37 & $652.0\pm2.7$ & $340.2\pm1.1$ &  $-1.0\pm0.1$ &  \\
   & C1-E$^{\rm b}$  & 20:21:28.82 & 40:48:46.31 & $10.0\pm1.5$ & $5.2\pm0.5$ & $-1.0\pm0.3$ & \\

\hline
 \end{tabular}
 \begin{tablenotes}
     \item[a] The Roman Squid flux density and the spectral index were calculated on the complete source and not as the sum of the four sub-components A, B, C, and D. 
     \item[b] Sources A, B, C, G78.4+2.6, C2, and YSO were catalogued in \citet{Benaglia2020} as BIC610-0666, BIC610-0665, BIC610-0672, BIC610-0676, BIC610-0737, and BIC610-0808, respectively. Sources C1-W and C1-E were catalogued as BIC610-0895 and BIC610-0916, and sources C1-MN and C1-MS were catalogued as a unique source, BIC610-0898.
     \item[c] Ultra compact H\,II region called VLA\,1 by \citet{Neria2010}.
     \item[d] Hard X-ray source \citep{Uchiyama2002, Leahy2013}.
   \end{tablenotes}
  \end{threeparttable}%
  \label{tab:results}
\end{table*}

\subsection{The ultracompact H\,II region G78.4+2.6 (\#2)}

\begin{figure}
   \centering
   \includegraphics[width=9cm, angle=0]{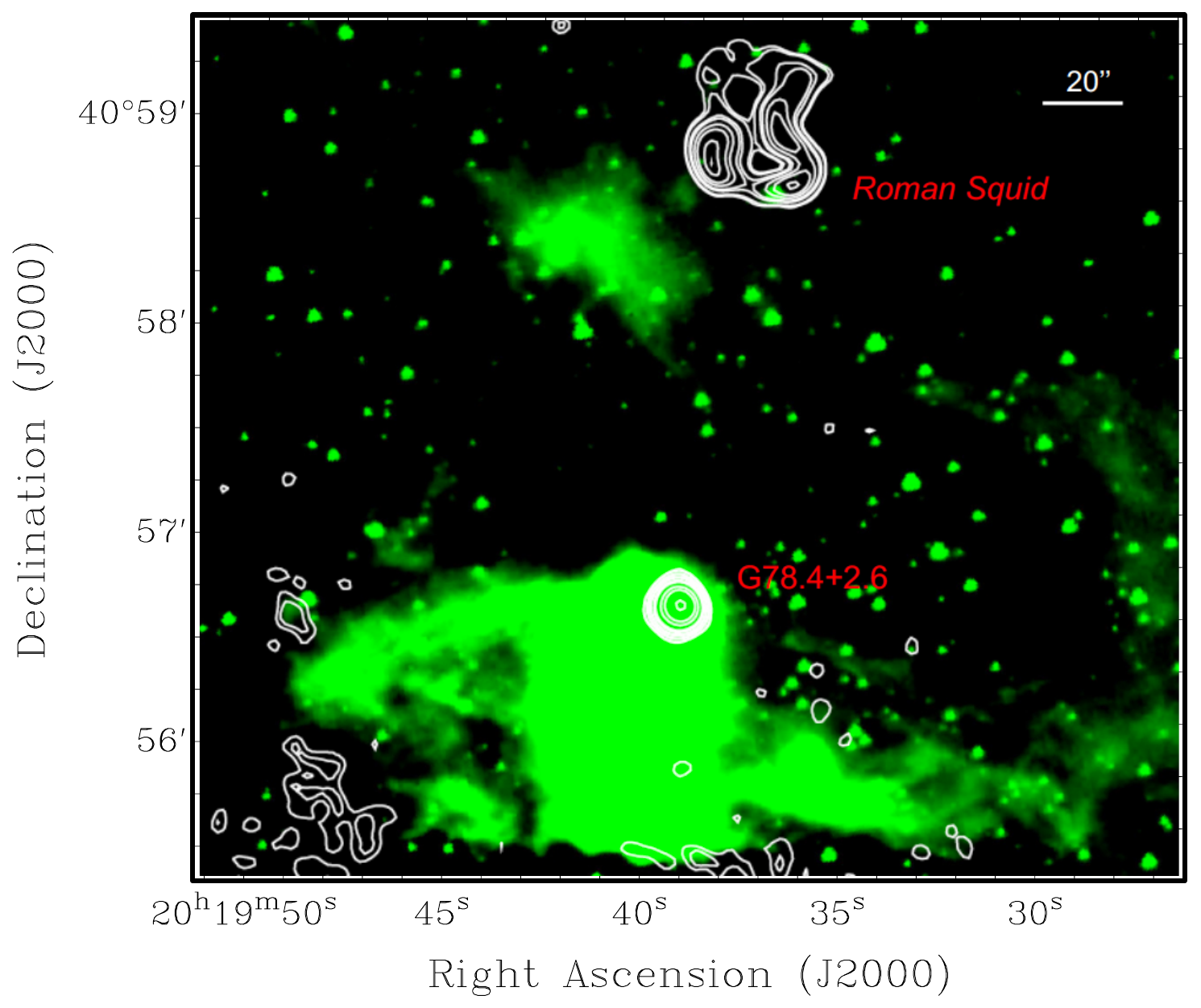}
      \caption{\footnotesize Spitzer 3.6~$\mu$m band emission image (green colour) of the region around the ultracompact H\,II region G78.4+2.6. The GMRT 610 MHz radio emission of this area is shown in white contours, which are 3, 4, 8, 10, 15, 20, 30, 35, 40, 80, 120, and 240 times the rms noise (0.13~mJy~beam$^{-1}$). The source Roman Squid is visible to the north.
              }
         \label{FigHII}
   \end{figure}

IRAS 20178+4046 (G78.4+2.6) is a massive star-forming region that has been classified as an ultra compact H\,II region \citep{Kurtz1994}. It has a cometary morphology in the infrared (see Fig.~\ref{FigHII}, green colour), and it is placed at an estimated distance of 1.7~kpc (see \citealt{Neria2010} and references therein).  Observations carried out with the VLA at 3.6~cm (8.4~GHz) wavelength detected an extended source, called VLA-1 \citep{Neria2010}, with a flux density of $62.26\pm0.14$~mJy and in positional coincidence with G78.4+2.6 \citep{Neria2010}. This value of the flux density is in agreement with values obtained in the 0.6--8.4~GHz frequency range from other authors \citep{McCutcheon1991,Kurtz1994,Tej2007}, and it is consistent with an optically thin region. The 8.4~GHz observations also detected four other fainter sources around G78.4+2.6, all of them having a flux density lower than 0.5 mJy. VLA\,1 has 2MASS 20193932+4056358 as a near-infrared counterpart.

Our GMRT observations  detected the source at both frequencies, showing a spherical morphology resolved at both frequencies. We remind readers that the synthesised beams   were $10'' \times 10''$ for the 325 MHz band, and $6'' \times 6''$ at the 610 MHz band. In Fig.~\ref{FigHII}, the GMRT 610 MHz data are represented in white contours, whereas the Spitzer data are represented in green. The UC~H\,II region is at the bottom of the image, showing the radio data (white contours) in positional coincidence with the region with the highest infrared emission. The Roman Squid is visible to the north of this region.
We obtained a flux density value of $36.5\pm1.4$~mJy at 325~MHz and of $63.9\pm0.8$~mJy at 610~MHz, which give us a spectral index of $\alpha=0.9\pm0.1$. The size of the source at 325 MHz is $28.1''$, whereas at 610 MHz the source has a slight elongation in the east-west direction and a size of $\sim 19.4''$.
Our flux density value at 610~MHz is in agreement with other radio estimates at the same frequency or higher, indicating, as mentioned before, an optically thin region. However, our lower flux density obtained at 325~MHz suggests that at this frequency we are observing the optically thick thermal region and that the spectrum has a turnover around 610~MHz. This turnover at low frequencies, and the flat spectrum at higher frequencies, is in agreement with the expected spectrum of an H\,II region.
Usually, H\,II regions are dominated by thermal emission. However, multi-wavelength observations have in some cases shown the presence of non-thermal emission. This non-thermal emission (synchrotron radiation) can be produced through the acceleration of thermal electrons up to relativistic energies in H\,II regions through the first-order Fermi acceleration mechanism \citep{Padovani2019}. However, the lack of non-thermal emission in the H\,II region G78.4+2.6, suggesting the absence of relativistic particles, makes us discard this source as a plausible gamma-ray contributor.

\subsection{Regions of enhanced radio emission: R1 and R2}

There are several areas with enhanced diffuse radio emission in the north-west region of the $\gamma$\,Cygni SNR and its surroundings. Two of these regions, dubbed R1 and R2, are positionally near the peak emission of MAGIC\,J2019$+$408, and they have a flux density at 325 MHz of $2.0\pm0.4$~Jy and $6.5\pm1.5$~Jy, respectively. The other areas of enhanced diffuse emission are more distant from the TeV peak position, but still within the area of emission of MAGIC\,J2019$+$408 (see Fig.~\ref{RadioSNR} and Fig.~\ref{MAGIC-SNR}). R2 corresponds to the most prominent overlapping between the SNR and the MAGIC source. Its spectral energy distribution at 325 MHz is $\nu L^{\rm 325MHz}_\nu\approx 10^{31}$~erg~s$^{-1}$, with a distance of 1.8 kpc.
The flux density of regions with no enhancement and covering similar areas, for instance as R1, decays to a few hundreds of mJy.

One can explore the possibility that part of the radio emission at 325~MHz is produced by secondary $e^\pm$-pairs ($e^\pm$) produced in proton-proton ($pp$) interactions in the region of the MAGIC source. In \cite{MAGIC2020}, hadronic emission through $pp$ interactions was favoured as the most likely mechanism behind the very high-energy radiation in that source. From the relation between the properties of the different outcomes and the parent protons in optically thin $pp$ interactions \citep{kel06}, one would expect that around one half of the gamma-ray luminosity went into secondary $e^\pm$. That would mean that for the MAGIC source roughly overlapping with the R1 and R2 sources, the secondary luminosity budget would be $L_\gamma\sim 10^{34}$~erg~s$^{-1}$ at a 1.8~kpc distance. The cooling times from synchrotron, inverse Compton (IC), and Bremsstrahlung \citep{blumenthal70} are roughly  
$$t_{\rm sync}\approx 1.4\times 10^4(30\,{\rm \mu G}/B)^2(1\,{\rm GeV}/E_e)\,{\rm kyr}\,,$$ 
$$t_{\rm IC}\approx 3.1\times 10^4(10\,{\rm eV~cm}^{-3}/u_{\rm IR})(1\,{\rm GeV}/E_e)\,{\rm kyr}$$ 
and 
$$t_{\rm Br}\approx 1000\,(30\,{\rm cm}^{-3}/n_{\rm ISM})\,{\rm kyr}\,,$$
respectively, for $e^\pm$ emitting around 325~MHz. These are much longer than the SNR age of $t_{\rm SNR}\approx 7$~kyr, and thus the secondary $e^\pm$ keep their original injection energy dependence, which would be $\sim E^{-2}$ given the gamma-ray
spectrum. Again, given the much shorter SNR age, the secondary $e^\pm$ synchrotron luminosity expected from the MAGIC source can be estimated using the gamma-ray luminosity above as
$$L_{\rm radio}\sim (1/2)L_\gamma(t_{\rm SNR}/t_{\rm sync})\sim 5\times 10^{30}(t_{\rm SNR,7\,kyr}/t_{\rm sync,14000\,kyr})\,{\rm erg~s}^{-1}\,.$$
This formula is based on the fact that secondary $e^\pm$ accumulate in the region along the SNR lifetime as dictated by the adiabatic cooling timescale, which is roughly the SNR age. Thus, the synchrotron luminosity is about the secondary $e^\pm$ accumulated energy divided by the corresponding radiative cooling timescale (this approach is valid when non-radiative losses dominate). The derived estimate is for the whole secondary synchrotron emission and this radiation is broadband, and the actual value of the predicted $\nu L^{\rm 325MHz}_\nu$ should be smaller by at least a factor of a few. On the other hand, the predicted $\nu L^{\rm 325MHz}_\nu$ could also somewhat increase by adopting a higher magnetic field $B$ ($E_e$ is fixed by $B$ and the radio frequency of interest).

To summarise, although these rough estimates suggest that a non-negligible amount of 325~MHz radiation from sources R1 and R2 could originate from secondary $e^\pm$ produced by $pp$ interactions in the MAGIC source region, a pure secondary emission origin for R1 and R2 does not seem likely. It is worth mentioning that, given the involved timescales and the secondary luminosity budget, the expected secondary gamma-ray luminosity from Bremsstrahlung and IC emission would be somewhat lower and much lower, respectively, than that of the primary $pp$ gamma rays in the region of R1 and R2. Moreover, given the rather steep spectrum of the TeV emission, significant X-ray synchrotron emission is not expected, but a faint source cannot be ruled out. 

Given that the secondary $e^\pm$ emission can account for a significant fraction, but likely not all of the radio emission from these radio sources, an additional primary electron population would be required to reach the observed fluxes and its energetics should be (at least) similar to that of the secondary $e^\pm$ from
$pp$ interactions. As for protons, the normalisation of secondary $e^\pm$ is determined by $t_{\rm SNR}$, but their injection luminosity should be a bit less than $\sim 0.1$ that of protons; given the secondary emission levels predicted, the fraction of primaries should be similar (or somewhat larger). All this thus implies a somewhat high electron-to-proton ratio of $\sim 0.1$ in the radio sources. We note that higher $B$-values would reduce this ratio as $\propto B^{-1.5}$.

\section{Other radio sources in the field of $\gamma$\,Cygni SNR}

In the following we present other relevant sources that are outside the MAGIC\,J2019$+$408 boundaries. 

\subsection{A young stellar object (\#4)}

Using mid-infrared photometry, \cite{Kryukova2014} classified the source [KMH2014] J202036.15+405753.58 as a young stellar object (YSO). This source appears in the 1.4 GHz NRAO VLA Sky Survey (NVSS) \citep{Condon1998} as NVSS\,J202036+405754 with a flux density of 76.8$\pm$2.4 mJy and in the Westerbork Northern Sky Survey \citep{2000yCat.8062....0D} as WN\,B2018.8+4048, with a flux density of $41\pm11$ mJy at 325 MHz. The derived spectral index is $\alpha_{\rm 325-1400}=0.4\pm0.2$.

Our observations at 325 and 610~MHz show an unresolved structure at both frequencies that is positionally coincident with the YSO within errors, and with a flux density of $189.8\pm1.3$~mJy at 325~MHz and $227.0\pm0.4$~mJy at 610~MHz, which gives a spectral density index of $\alpha_{\rm 325-610}=0.3\pm0.1$. This spectral index is compatible within errors with the value derived at higher frequencies, $\alpha_{\rm 325-1400}=0.4\pm0.2$. However, the flux density at 325 MHz of WN\,B2018.8+4048 is a factor 4.5 lower than our GMRT results, apparently indicating that the source could be variable. This is also supported by the low value of 76.8$\pm$2.4 mJy obtained at 1.4 GHz by NVSS when compared to the value of 291.2 mJy at 1.4 GHz derived by extrapolating the GMRT data. Given the hard radio spectra found, the radio emission could be produced by a thermal mechanism (free-free) associated with a jet in the YSO \citep{Anglada2018}.

At X-rays the source has been detected by {\it Chandra} (2CXO\,J202036.2$+$405753) \citep{Evans2019} and {\it XMM-Newton} (4XMM\,J202036.3$+$405753) \citep{Webb2020} as a point source. The infrared counterpart (WISE\,J202036.18$+$405753.9, UGPS\,J202036.13$+$405753.7) is a star of magnitude {\it K}=16.4.

\subsection{Clumps C1(\#5) and C2 (\#3)}

X-ray observations with {\it ASCA} of the SNR showed several clumpy sources in the northern region in the hard energy band. These clumps were referred to as C1 and C2 \citep{Uchiyama2002}. 

\begin{figure}
   \centering
   \includegraphics[width=9cm, angle=0]{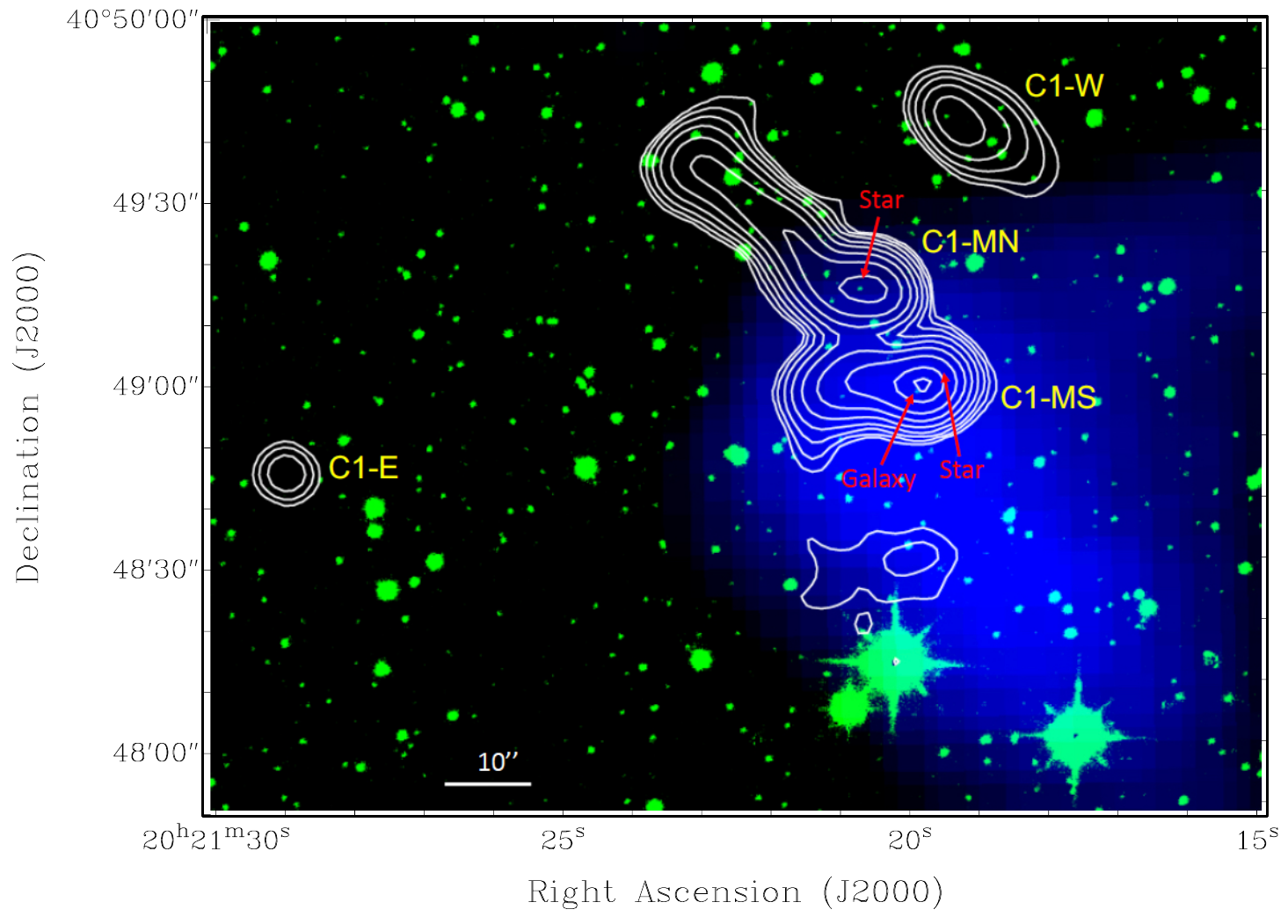}
      \caption{\footnotesize Archive {\it Chandra} image of the clump C1 smoothed with a 12-arcsec Gaussian function (blue) and the {\it K}-band image from UKIDSS (green). The overlaid white contours represent the 610 MHz radio emission from GMRT. Contour levels are 6, 12, 24, 48, 96, 192, 384, and 768 times the rms noise (0.13 mJy beam$^{-1}$). The northern part of the source C1-MN, and the sources C1-W and C1-E, fell just off the {\it Chandra} CCD chips.
              }
         \label{FigC1}
   \end{figure}

{\bf C1 clump}: The archive {\it Chandra} image of the region (PI Mark Theiling) shows that C1 is an extended hard X-ray source with a column density that indicates that it is an extragalactic background source \citep{Leahy2013}. Fig.~\ref{FigC1} shows the archive {\it Chandra} image of the clump C1 with the 610~MHz contours of the GMRT observations overlaid. The radio data show a cluster of sources, labelled C1-W, C1-MN, C1-MS, and C1-E. C1-MN is the brightest and largest of them, with a flux density of 939.0~mJy at 325~MHz and of 414.5~mJy at 610~MHz, and a jet of $\sim 26^{\prime\prime}$ pointing to the north-east. C1-MS has a shorter jet, pointing east to west. The core of C1-MN and C1-MS overlap with the X-ray clump, whereas the northern part of C1-MN falls just off the CCD chip. Also, the sources C1-W and C1-E fell off the CCD chips. 
It is possible that the X-ray emission could be extending to the north, making an association between C1-MN, C1-MS, and C1-W with the X-ray C1 clump possible. The value of the spectral index of all these three sources (see Table~\ref{tab:results}) indicates a non-thermal origin of the radio emission, compatible with being extragalactic objects, as indicated by the X-ray absorption, although the region is complex and sources of a different nature may be overlapping.
Using the GMRT 150 MHz all-sky radio survey \citep[TGSS ADR1\footnote{http//tgssadr.strw.leidenuniv.nl/docu.php, TIFR GMRT Sky Survey Alternative Data Release 1.}, ][]{Intema2017}, we have imaged the region around the C1 clump showing an unresolved elongated structure with an angular size of $21^{\prime\prime}\times 78^{\prime\prime}$ and with a flux density of 2340 mJy. This structure slightly overlaps with the sources C1-MN and C1-MS observed at the higher frequencies. These results support the non-thermal nature of the main structures in the C1 clump.

The UKIDSS K-band image of this area \citep{Lucas2008} shows different possible infrared counterparts of the C1 sources (see Fig.~\ref{FigC1}).  Carrying out a search of infrared sources in a radius of $3^{\prime\prime}$ around the peak emission of each C1 radio source, we have found a star (UGPS\,J202120.62+404914.8) with a magnitude of {\it K}=17.85 at $0.6^{\prime\prime}$ from the radio peak of C1-MN. In C1-MS, there is a galaxy (UGPS\,J202119.80+404858.5) of {\it K}=16.87 placed at  $1.0^{\prime\prime}$ from its radio peak and a star (UGPS\,J202119.65+404900.2), classified in the catalogue with a probability between 70 and 90\% at $1.3^{\prime\prime}$ and a magnitude of {\it K}=18.2. In C1-W, there are three stars within a radius of $3^{\prime\prime}$ from the radio peak and with magnitudes {\it K}=18.7, 16.9, and 14.8.

{\bf C2 clump}: The ASCA observations of the C2 clump, centred on RA$_{\rm J2000}$ = 20h20m00s, DEC$_{\rm J2000}$ = $40^{\circ}40'03''$, showed that the source has an extended structure of a few arcminutes \citep{Uchiyama2002}. Images by {\it Chandra}, with a higher sensitivity and spatial resolution, showed that its emission is dominated by a point source at RA$_{\rm J2000}$ = 20h20m12s, DEC$_{\rm J2000}$ = $40^{\circ}39'00''$, rather than an extended source \citep{Leahy2013}. This source can be fitted with a pure power-law spectrum and a column density consistent with the $\gamma$\,Cygni SNR, although the large errors on $N_{\rm H}$ prevent one from definitively saying  whether or not C2 is associated with the $\gamma$\,Cygni SNR \citep{Leahy2013}. \citet{Bykov2004} also reported a bright hard X-ray (25–40 keV) clump, which is nearly coincident with the {\it ASCA} C2 source. 

\begin{figure}
   \centering
   \includegraphics[width=9cm]{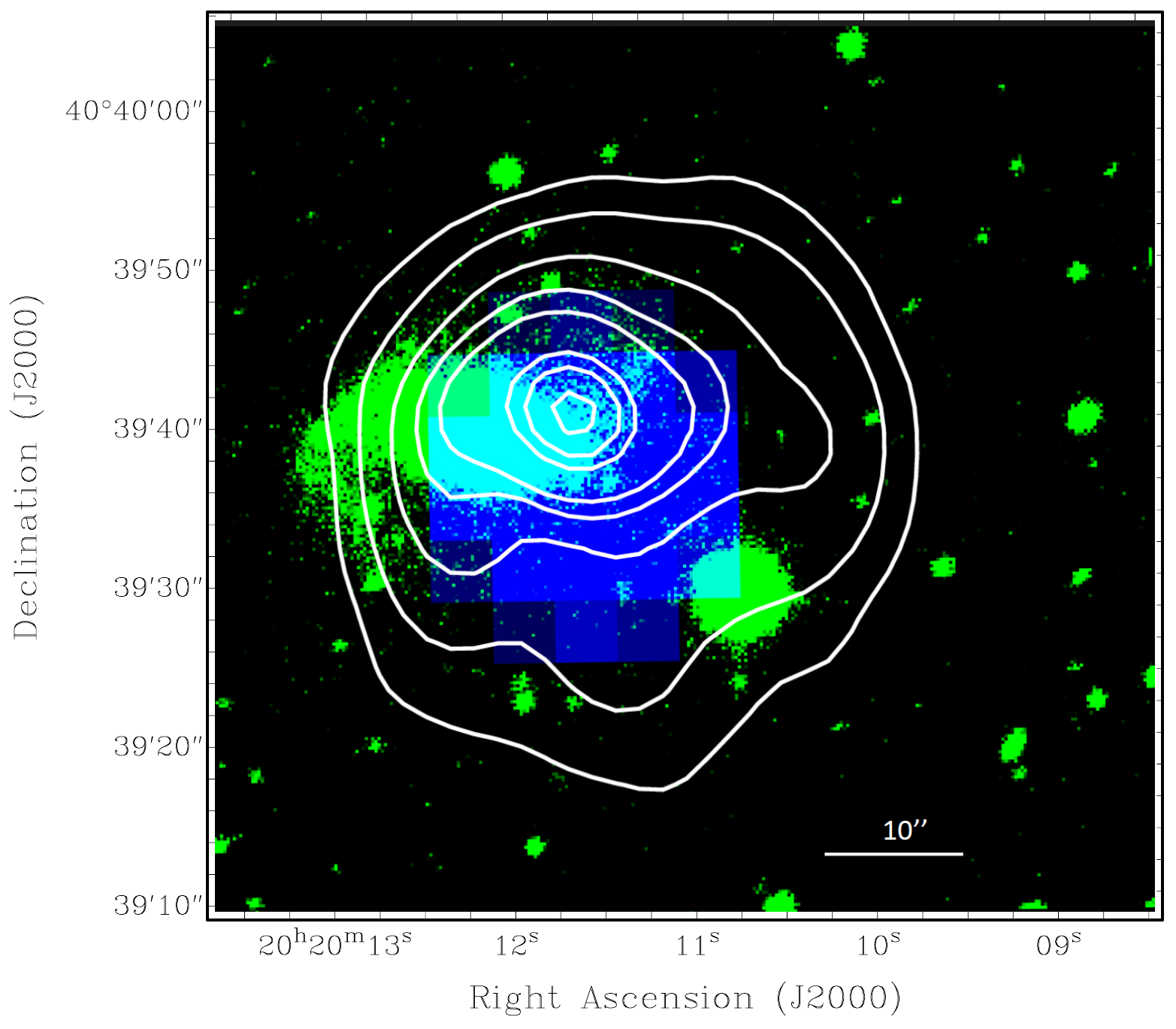}
      \caption{\footnotesize {\it K}-band image from UKIDSS (green) and archive {\it Chandra} image of the clump C2 (blue). The overlaid white contours represent the 610 MHz radio emission from GMRT. Contour levels are 5, 10, 20, 30, 40, 80, 100, and 130 times the rms noise (0.25 mJy beam$^{-1}$). The {\it K}-band image clearly shows a galaxy with a size of $\sim23"$, which is positionally coincident with the radio and X-ray emission.
              }
         \label{FigC2}
   \end{figure}

Our radio observations of this region show a bright source, labelled \#3 in Fig.~\ref{RadioSNR}, which is positionally coincident with the hard X-ray source detected by {\it Chandra} \citep{Leahy2013}. We maintain the name of C2 for this
source (see Table~\ref{tab:results}). The shape of the source is nearly circular, with a slight elongation from east to west in the core of the source and with a size of $\sim 40^{\prime\prime}$ at 325 and 610~MHz. 
The source C2 has a flux density of $403.9\pm12.4$ mJy at 325 MHz and of $167.1\pm3.2$ mJy at 610 MHz, implying a spectral index of $-1.4\pm0.1$. In Fig.~\ref{FigC2} we show a composite image of the C2 region observed at 610 MHz together with an infrared and X-ray image. The UKIDSS {\it K}-band image (in green) \citep{Lucas2008} clearly shows a galaxy (UGPS\,J202011.56+403938.1) that overlaps with the {\it Chandra} X-ray emission (in blue) and with the GMRT radio emission (white contours). 

%





\section{Conclusions}

In what follows, we present the main conclusions of this work about the different sources that were investigated. First, we start with the GMRT 610~MHz images of the Roman Squid and the C1 clump, which have shown a complex structure in both sources.

In the case of the radio morphology of the Roman Squid, the existence of a galaxy at the south of the structure suggests an extragalactic origin for this source, although a galactic origin cannot be ruled out. To explore the head-tail scenario, it is mandatory to make deeper radio observations at different frequencies to see if the expected radio structure (tail) extending further north for several arcminutes is present. In addition, X-ray observations could be able to provide crucial information about the morphology of the Roman Squid in the X-ray band and, ultimately, using spectral information, the distance to the source. 

Regarding the C1 clump, our GMRT 610~MHz observations have also resolved this source in several components that appear to be independent, although all of them have a spectral index of $\sim-1$ in common. One of these components, C1-MS, overlaps with an X-ray source of extragalactic origin, and its radio core has a galaxy as a counterpart. All of this makes C1-MS a likely extragalactic source. Another component, C1-MN, is the strongest component and at the same time has a long jet. Near the core of this source, there is a star, and if both C1-MN and the star were physically associated, then the former could be a microquasar \citep{Paredes2011}.

In the case of the source associated to the C2 clump, the X-ray column density is consistent with that of the $\gamma$\,Cygni SNR \citep{Leahy2013}. However, the presence of a galaxy aligned with the radio and X-ray counterparts seems to indicate an extragalactic origin for this source, assuming that the infrared, radio and X-ray emission all come from the same source.

Finally, the extended radio sources R1 and R2 seem to correspond to radio counterparts of the TeV source detected by MAGIC. Our estimates show that a significant fraction of the 325~MHz emission from these radio sources may come from synchrotron radiation of secondary $e^\pm$, produced locally by $pp$ interactions in dense regions of the ISM bordering the northern edge of the SNR. However, given the expected energetics of the secondary $e^\pm$, a pure secondary origin for the observed radio emission does not seem probable, although more detailed modelling is necessary to better assess the contribution of these $e^\pm$ to the emission from the region detected in radio, and potentially at higher energies. It is noteworthy that the levels of radio emission from these sources imply that the electron-to-proton ratio in the region, assuming $pp$ interactions as the origin of the gamma rays, may be $\sim 0.1$, distributed between secondary $e^\pm$ and a primary electron population.

\begin{acknowledgements}
     We thank the staff of the GMRT that made these observations possible. GMRT is run by the National Centre
for Radio Astrophysics of the Tata Institute of Fundamental Research. PB acknowledges support from ANPCyT PICT 0773-2017. CHIC acknowledges the support of the
Department of Atomic Energy, Government of India, under the project 12-R\&D-TFR-5.02-0700. JMP and VBR acknowledge
financial support from the Spanish Agencia Estatal de Investigaci\'on through the grant PID2019-105510GB-C31/ AEI / 10.13039/501100011033 and `Unit of Excellence María de Maeztu 2020-2023' award to the Institute of Cosmos Sciences (CEX2019-000918-M).
JMP and VBR are correspondent investigators of CONICET, hosted at IAR.
This research has
made use of the SIMBAD and Vizier databases, operated at CDS, Strasbourg, France, and of NASA's Astrophysics
Data System bibliographic services. 
\end{acknowledgements}

%
%



\bibliographystyle{aa}
\bibliography{references.bib}{}

\end{document}